\documentclass[twocolumn,preprintnumbers,amsmath,amssymb]{revtex4} 
\usepackage{graphicx}
\usepackage{dcolumn}
\usepackage{bm}

\begin{document} 
\title{On the nature of interlayer interactions in a system of two graphene 
fragments} 
\author{Julia Berashevich and Tapash Chakraborty} 
\affiliation{Department of Physics and Astronomy, University 
of Manitoba, Winnipeg, Canada, R3T 2N2} 
\begin{abstract} 
With the help of the quantum chemistry methods we have investigated the 
nature of interlayer interactions between graphene fragments in different 
stacking arrangements (AA and AB). We found that the AB stacking pattern 
as the ground state of the system, is characterized by the effective 
inter-band orbital interactions which are barely present in the AA. Their 
vanishing induces electronic decoupling between the graphene layers,  
so that the bonding interaction $\Delta E_{oi}$ between the flakes is 
drastically reduced from $-0.482$ eV to $-0.087$ eV as the stacking pattern 
is changed from AB to AA. The effective way to improve the bonding interaction 
between layers preserving the same AA lattice order is to induce rotation of 
the layer. As the flake is rotated, the bonding interactions are 
improved mostly due to suppression of the Pauli repulsion which in turn 
increases the interlayer orbital interactions, while the inter-band 
part of those remain negligible on the whole range of the rotation angle. 
The Pauli repulsion is also found to be the main force moving apart two 
fragments as the stacking pattern is changed from AA to AB. This enhances 
the equilibrium interlayer distance, which for the AA staking is larger than 
the established value for the AB stacking (3.4 \AA). 
\end{abstract} 
 
\maketitle 
Bilayer graphene \cite{varchon,naturp,hass,reina,warner,decouplt,mccan}, 
being regarded as an important system for applications in semiconductor 
electronics has intrigued the scientific community to look deeper into 
the nature of interlayer interactions in this system. Graphene layers stacked 
together is uniquely different from other solid state materials because of 
the interlayer weak van der Waals and steric interactions, 
instead of the occurrence of C-C bonding between the layers. 
The half-filled $p$ orbital left on each carbon 
atom after bonding with its neighbors within the honeycomb lattice is responsible 
for formation of the $\pi$ bonds between two neighboring atoms within the same 
layer. This creates a closed shell electron system carrying the weakly bound 
$\pi$ electrons that are distinct due to their high mobility within the graphene 
layer \cite{mob,review,review1}. Stacking two systems of closed electron shells 
would cause the interlayer interactions to be mostly repulsive which results 
in the expulsion of valence electrons from the overlap region such that 
only a weak electronic coupling between the layers can occur. If under certain 
conditions the electronic coupling becomes negligible then each layer would 
display its own electronic behavior in the band diagram as was found recently 
in twisted bilayer graphene \cite{latil,shall,lopes}. 
 
Even before the discovery of graphene the interlayer interactions in natural 
graphite, which is basically a system of stacked graphene layers, received 
intense attention \cite{types,yosh,ruuska,charlier} where the effect of 
interlayer decoupling was not encountered. Three types of layer arrangements 
are known to exist in graphite \cite{types}, but the most common one is 
the Bernal stacking in which the carbon atoms belonging to different  
sublattices A and B form the AB stacking pattern between the layers. Contrary 
to the conventional wisdom that only the long-range van der Waals interaction
is important in the case of stacking of closed shell systems, for the AB stacking 
in graphite it was shown that the orbital overlap between the $\pi$ orbitals 
belonging to different layers \cite{yosh,our} is as essential as the all important 
van der Waals forces. Therefore, it was predicted that despite the well known 
interlayer distance of 3.4 \AA\ in natural graphite, for a system containing  
only two or three layers, the interlayer spacing depends on the number 
of layers \cite{yosh} due to different nodal interactions of the overlapping 
$\pi$ orbitals. For an odd number of adjacent layers the equilibrium spacing 
between the layers was predicted to be 3.30 \AA, while for the even number 
it is 3.58 \AA\ \cite{yosh}. It then clearly follows that in graphite the 
electronic coupling between the layers can not be neglected. Rather, it 
provides a substantial influence on the interlayer interactions in addition 
to the van der Walls and steric type of interactions.  
 
In this context, it is worth pondering what is acually happening with the 
interlayer interactions in twisted bilayer graphene where interestingly, one 
observes electronic decoupling between the layers. The decoupling was first  
observed experimentally \cite{varchon,naturp,hass} and was later investigated 
theoretically \cite{latil,shall,lopes}. Theoretical interpretations 
\cite{latil, shall,lopes} relate the decoupling to the occurrence of a misorientation 
of 2$^{\circ}$-5$^{\circ}$ between the layers. According to a proposed 
model, the layer rotation in the real space induces a displacement of the Dirac 
cones generated in each layer in the reciprocal space \cite{latil, shall,lopes} 
thereby causing the interlayer decoupling. Experimentally, the misorientation 
of 2$^{\circ}$-5$^{\circ}$ in AA-stacked bilayer graphene has been detected 
in systems that were created using various fabrication techniques such as the 
epitaxial growth \cite{hass,naturp,varchon}, chemical vapor deposition \cite{reina} 
and ultrasonication \cite{warner}. Appearance of the rotational misorientation that is independent 
of the fabrication techniques suggests the presence of some forces between 
the layers strong enough to cause the layer rotation. However, the available 
theoretical models \cite{latil,shall,lopes} on the electronic properties of the
AA stacked graphene deal only with the band properties but do not shed any 
light on the underlying physical reasons involved in decoupling, such as the 
interlayer forces.  
 
In a recent work \cite{our} we related the origin of the decoupling phenomenon 
and rotational misorientation with layer stacking pattern which is AA in fabricated 
multilayer graphene \cite{varchon,naturp,hass,reina,warner} against the AB stacking 
in natural graphite. For the AA staking, the interlayer electronic coupling is 
suppressed by a significant repulsion arising between the graphene layers \cite{our}. 
This repulsion is also expected to be responsible for the occurrence of lattice 
misorientation between the layers. It was suggested that rotational misorientation, which 
creates the Moir\'e pattern, appears as a way to suppress the repulsion, thereby 
lowering the total energy of the system. Even a slight layer rotation 
of $\sim$2$^{\circ}$-5$^{\circ}$ substantially shrinks the areas characterized 
by the AA lattice superposition in which repulsion dominates over other forces (the 
areas with AA and AB stacking coexist in the Moir\'e pattern). The other important 
result was a prediction \cite{our} that the strong repulsion may induce bumps on 
the graphene surface in the areas where AA stacking is preserved. All these effects 
are important and require careful studies because the phenomenon of layer rotation 
through electronic coupling between layers can offer ways to manipulate the 
electronic properties of twisted graphene (Moir\'e pattern of different rotation 
angle is characterized by different percentage of AA-spotted areas). Therefore, 
in this work we present a detailed quantitative analysis of the repulsive forces 
and the orbital overlap in stacked graphene layers and their alteration with the 
appearance of rotation. Our studies are based on the density functional methods 
including a recently proposed empirical correction (Grimme correction \cite{grimme}) 
which was developed for a proper consideration of the dispersive interactions 
between the closed shell electron systems.  
 
\section{Computational methods} 
The computations were performed with the ADF quantum chemistry code \cite{adf} 
which uses the Kohn-Sham approach to density functional theory (DFT). The 
Kohn-Sham approach replaces the many-body system within the Hamiltonian equation 
by a system of the non-interacting particles while all the many body terms are 
incorporated into the so-called Kohn-Sham potential. This concept is quite useful 
in the investigation of interacting closed shell systems because it allows us to 
present each graphene flake as an isolated fragment and two fragments interacts as
the flakes are stacked. In this way, a proper investigation of the forces and the 
orbital overlap can be performed directly in terms of the fragment presentation.  
 
Within the ADF code the forces between fragments are included in the bonding energy 
$\Delta E^0$ which comprises of several majors components \cite{bick} ($\Delta E^0$=
$\Delta V_{el}$+$\Delta E_{p}$+$\Delta E_{prep}$+$\Delta E_{oi}$+$E_{dis}$). The 
first component ($\Delta V_{el}$) takes care of the interactions of electrostatic 
nature related to the modification of the charge distribution (originated from the 
charge transfer between occupied and unoccupied orbitals), when two systems are 
allowed to interact. The second one is the energy change induced by the Pauli 
repulsion ($\Delta E_{p}$), which include several components; exchange repulsion, 
kinetic repulsion, overlap repulsion, all results from obeying the Pauli antisymmetry 
principle. The next term $\Delta E_{prep}$ describes the energy required to change 
the conformation of the fragments (structural modification) from the initial geometry 
containing separate fragments to the final geometry where the fragments are allowed to 
interact. The bonding interactions between two fragments are included in the $\Delta 
E_{oi}$ term which originates from the overlap of the fragment's orbitals. The last 
term $E_{dis}$ is the empirical dispersion correction introduced by Grimme \cite{grimme} 
and its magnitude is defined by the long-range van der Waals interactions, whose 
contribution in the short range is reduced by the damping function. 
 
Even though the $\Delta E_{oi}$ term is a measure of the orbital overlap, the 
interlayer forces such as the Pauli repulsion and orbital polarization contribute 
to the $\Delta E_{oi}$ as well. The effect of interlayer forces can not be discarded 
from $\Delta E_{oi}$ and so the overlap of the selected orbitals can not be separated 
from the others. This makes it hard to get a proper understanding of the intricacies 
of interlayer interactions between two fragments. The most effective way to proceed 
is to follow the established method of linear combinations of the orbitals. This method 
can be applied for the results obtained with the ADF program. With the ADF, each fragment 
is described by its own set of orbitals and the program facilitates their mixing upon 
the inclusion of the interaction between the fragments. Therefore, the fragment approach 
allows us to evaluate the overlap matrix $S_{i,j}$ between the fragments $i$ and $j$ 
of the Kohn-Sham Hamiltonian ($\langle \varphi^{}_i|h^{}_{\rm KS}|\varphi^{}_j\rangle$) 
directly in terms of the linear combinations of the fragment orbitals via the relation 
$h_{\rm KS}=SCEC^{-1}$, where $C$ is the eigenvector defined in terms of the fragment orbitals 
and $E$ is the eigenvalue matrix \cite{senthil}. The overlap matrix $S$ purely depends on the form 
of the interacting orbitals and on the distance that keeps the two fragments apart neglecting 
the contribution from the attractive and repulsive forces arising between the fragments. We used 
the overlap matrix $S_{i,j}$ to define the spatial overlap integral between the fragments $i$ 
and $j$, which is $J_{i,j}=\langle \varphi^{}_i|H|\varphi^{}_j\rangle$. 
 
In this work we consider the spatial overlap integral $J_{i,j}^{H-H}$ between the highest occupied 
fragment orbitals (HOFO), i.e., between two $\pi$ orbitals, each located on different fragments  
while their overlap defines the HOMO of the joint system. The overlap integral was also 
calculated between the $\pi$ and $\pi^*$ orbitals, i.e., between the highest occupied orbital of 
one fragment (HOFO$_{1}$) with the lowest unoccupied orbital of another fragment (LUFO$_{2}$)  
and because there are two parts of such interactions, HOFO$_{1}$-LUFO$_{2}$ and HOFO$_{2}$-LUFO$_{1}$, 
the average value of overlap integral was considered and combined into the $J_{i,j}^{H-L}$. 
 
We used the hybrid BLYP exchange-correlation functional, applying the empirical dispersion 
correction 1.05 recommended by Grimme \cite{grimme,grisha}. For the interacting molecules of 
closed electron shells it was found that the proposed correction is enough to reproduce the 
intermolecular distance to what is observed in the experiments or achieved with a more accurate level 
such as the {\it ab initio} M\o ller-Plesset second-order (MP2) method \cite{grimme}. For a proper 
description of the tails of the electron wavefunctions that is important for long-range interactions,  
we used the Slater-type orbitals. The quite extended TZP basis set (triple-$\zeta$ polarized basis 
set) was applied in all the calculations which improves the precision of our results while suppressing 
the basis set superposition error \cite{hobza}. We tested the chosen method to reproduce the 
interlayer distance between the graphene flakes stacked in the AB pattern (which is well known to 
be 3.4 \AA\ in natural graphite) and indeed the correct interlayer distance was obtained (the so-called 
equilibrium distance of the AB pattern $d_{eq(AB)}$). For this calculation the atomic coordinates 
within the graphene plane (along the $x,y$ directions) were confined and only the coordinates in 
the $z$ direction, i.e., perpendicular to the graphene plane, were used for relaxation. In fact, the 
full relaxation of the system of two stacked fragments is problematic as 
the repulsion between layers leads to sliding of the fragments away from each other. 
 
Our main emphases in this work are the energy decomposition analysis of the bonding energy 
$\Delta E^0$ ($\Delta V_{el}$, $\Delta E_{p}$, $\Delta E_{prep}$ and $\Delta E_{oi}$) 
and the spatial overlap integrals ($J_{i,j}^{H-H},J_{i,j}^{H-L}$) which are determined by the 
stacking pattern between two graphene flakes (flake rotation) and the interlayer distance. In 
most cases the single point calculations have been used for which the contribution of 
$\Delta E_{prep}$ becomes zero.  
 
\section{AA and AB stacking} 
\begin{figure} 
\includegraphics[scale=0.41]{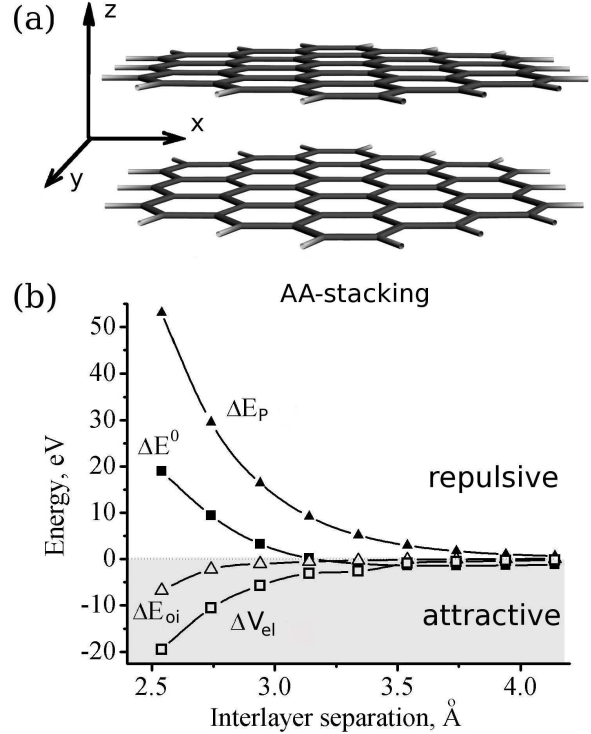} 
\caption{ (a) Stacked graphene flakes in the 3D space coordinate system. 
(b) The energy components ($\Delta V_{el}$, $\Delta E_{p}$,$\Delta E_{oi}$) 
of the bonding energy ($\Delta E^0$) as a function of the interlayer distance 
between two graphene fragments. $\Delta V_{el}$ is the electrostatic 
interactions, $\Delta E_{p}$ is Pauli repulsion, $\Delta E_{oi}$ is the orbital 
interactions energy.} 
\label{fig:fig1} 
\end{figure}

For our investigations we used two graphene flakes with the carbon atoms at the edges terminated  
by the hydrogen atoms, as shown in Fig.~\ref{fig:fig1} (a) with our goal to minimize the contribution  
of the localized states into the simulation results. Because bilayer graphene obtained in the 
experiments \cite{varchon,naturp,hass} has shown the AA stacking pattern instead of the AB,
common in natural graphite, we probe the AA stacking for the equilibrium distance between the flakes 
($d_{eq(AA)}$). It was found that this distance is indeed enhanced in the AA stacking up to 
$d_{eq(AA)}$=3.67 \AA\ against the $d_{eq(AB)}$=3.4 \AA\ for the AB stacking. To explain an increase in 
the interlayer distance we applied the decomposition analysis of the bonding energy.  
 
We performed the single point calculations in which the layers were separated by 
the established equilibrium distances ($d_{eq}$) found to be different for AA and 
AB stacking. The obtained interlayer forces and spatial overlap integrals are 
collected in Table ~\ref{tab:table1}. For the purpose of comparison, in addition to 
the lattice arrangements AA and AB, we also carried out the bonding analysis 
for the AA$^{\prime}$ stacking pattern for which the interlayer equilibrium distance 
was found to be 3.41 \AA\ and those results are also enclosed in Table ~\ref{tab:table1}.
The AB-conformation is characterized by a much stronger bonding interaction than that 
of AA, as defined by a more negative bonding energy $\Delta E^0$, thereby making the 
AB-stacking the ground state of the system. The structural distinction of the AB 
configuration from AA consists in sliding of one graphene flake relative to the other 
by 1.42 \AA\ along the $x$ axis that induces the shift of the bond positions between 
the layers against their superposition for the AA stacking (for the AA case the atomic 
coordinates are matched in the $x$ and $y$ directions for both layers). To obtain the 
AA$^{\prime}$ stacking a sliding of 1.23 \AA\ is applied along the $y$ axis.

For the AA stacking, due to the lattice superposition the $\pi$ clouds between the 
layers are also superposed that leads to their effective overlap defined by the spatial 
overlap integral $J_{i,j}^{H-H}$=0.443 eV. The flake sliding induced for the AB and 
AA$^{\prime}$ stacking breaks the bond superposition condition recognized for the AA 
stacking thereby inducing the disarrangement of the $\pi$ clouds within the overlapping 
region. As a result, the $\pi-\pi$ interlayer interactions, which is of particular 
interest since it is supposedly responsible for the occurrence of electronic coupling 
between the graphene layers \cite{mcdonald,mccan}, is significantly reduced for the 
AB stacking pattern. This is reflected by a suppression of the spatial overlap integral 
to $J_{i,j}^{H-H}$=$-$0.229 eV (even more drastic reduction is observed for the AA$^{\prime}$ 
case where $J_{i,j}^{H-H}$=0.023 eV). 
 
\begin{table} 
\caption{\label{tab:table1} The electronic properties and the interlayer forces between two 
graphene fragments stacked in AA ($d_{eq(AA)}$=3.67 \AA), in AB ($d_{eq(AB)}$=3.4 \AA) 
and in AA$^{\prime}$($d_{eq(AA)}$=3.41 \AA) configurations. All values are in eV.} 
\begin{tabular}{l|c|c|c|c|c|c|c|c} 
\hline 
& HOMO & LUMO & $J_{i,j}^{H-H}$ & $J_{i,j}^{H-L}$ & $\Delta E_{oi}$ & $\Delta V_{el}$  
& $\Delta E_{p}$ & $\Delta E^0$ \\ 
\hline 
AA & -4.024 & -2.596 & 0.443 & $10^{-5}$ & -0.087 & -0.691 & 2.136 & -1.483 \\ 
AB & -4.164 & -2.506 & -0.229 & -0.169 & -0.482 & -1.184 & 3.388 & -1.931 \\ 
AA$^{\prime}$ & -4.070 & -2.565 & 0.023 & -0.191 & -0.491 & -1.263 & 3.644 & -1.863 \\ 
\hline 
\end{tabular} 
\end{table} 

When two fragments are stacked, the majority of the orbital interactions described by 
$\Delta E_{oi}$ arise from the overlap of the $\pi$ orbitals ($\pi$-$\pi$ or $\pi$-$\pi^*$).
For interaction of two closed shell systems, the $\pi$-$\pi$ overlap is not the one 
that leads to the bonding interactions, and therefore, might be ignored within the 
orbital interaction term $\Delta E_{oi}$. That explains the contradictory behavior 
of the overlap integral $J_{i,j}^{H-H}$ and the $\Delta E_{oi}$ term, such that when 
a reduction of the overlap integral $J_{i,j}^{H-H}$ occurs for the AB stacking,
the orbital interactions $\Delta E_{oi}$ between the fragments is improved.
However, the improvement of $\Delta E_{oi}$ with modification of the lattice arrangement 
from AA to AB is also not consistent with behavior of the Pauli repulsion $\Delta E_{p}$ 
whose increase supposedly suppresses the interlayer orbital interaction $\Delta E_{oi}$.
Therefore, to understand the alteration of the $\Delta E_{oi}$ term we should take into 
consideration other components of the orbital interactions, such as the orbital polarization 
and the interlayer interaction of the $\pi$-$\pi^*$ orbitals.

\begin{figure*} 
\includegraphics[scale=0.90]{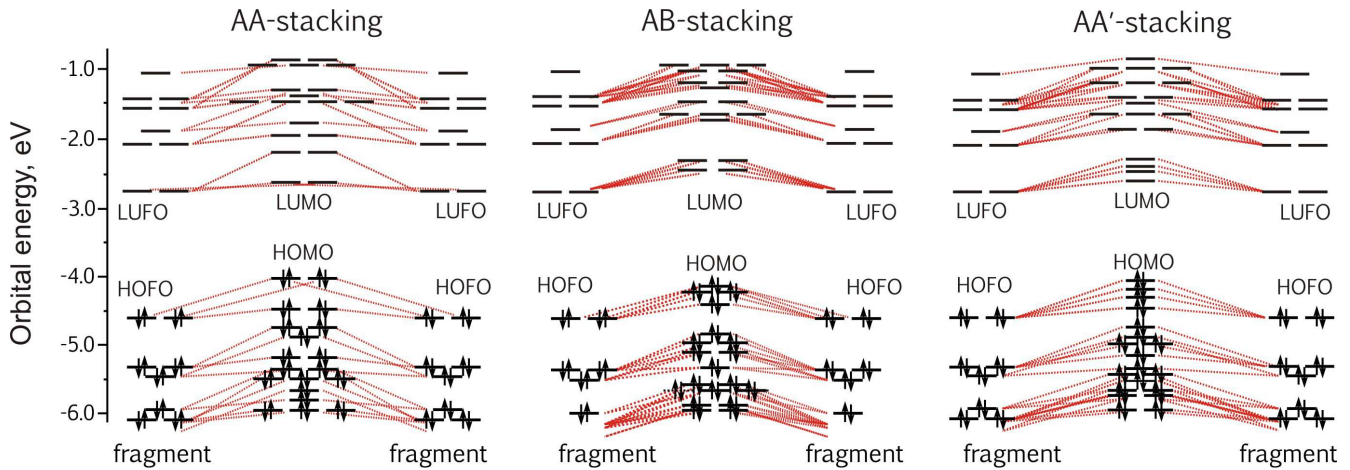} 
\caption{The energy diagram demonstrating the formation of the molecular orbitals 
(HOMOs and LUMOs) as the fragment's orbitals overlap (HOFO and LUFO) for different 
arrangements of graphene lattices in the case when the fragments are separated by the
equilibrium distance $d_{eq}$.} 
\label{fig:fig2} 
\end{figure*}

Orbital polarization reflects a mixing of the occupied/virtual orbitals in one fragment due to 
the presence of another fragment, i.e., each valence electron of one fragment entering the 
electron space of electrons of other fragment polarizes its orbitals. The polarization effect 
is caused by the repulsion arising between interacting electrons \cite{bills}. 
Analyzing the orbital formation after perturbation of the fragment's orbitals, 
a discrepancy in the product orbitals for the AB, AA and AA$^{\prime}$ stacking patterns
was detected as demonstrated by the scheme in Fig.~\ref{fig:fig2}. Let us consider 
the formation of the orbitals generating the HOMO-LUMO gap in the joint system which 
would give the most contribution into the interlayer interaction of the $\pi$-$\pi^*$ 
orbitals. In formation of HOMO orbital (LUMO) of the joint system two degenerate 
fragment's orbitals HOFO participates (two LUFO orbitals for the LUMO formation). 
Perturbation of those degenerate HOFO orbitals (overall four in two fragments) creates the
four molecular orbitals in the bilayer system (HOMO, HOMO$-$1, HOMO$-$2 and HOMO$-$3).
Similarly, the LUFO orbitals perturb in the valence band so that two LUFOs are taken 
from each fragment and their perturbation leads to formation of four LUMO orbitals in 
the stacked system (LUMO, LUMO+1, LUMO+2, LUMO+3). For the AA stacking, two pairs of 
product orbitals possess an identical orbital energy, i.e., HOMO and HOMO$-$1 (LUMO and 
LUMO+1) are degenerate, the same for HOMO$-$2 and HOMO$-$3 (LUMO+2 and LUMO+3). However, 
already for the AB case the conduction band is limited by a single HOMO orbital being 
a product of the perturbation of all four fragment's orbitals while the rest of the 
generated orbitals are shifted deeper into the conduction band where two of them still 
would remain degenerate. Mixing of the LUFOs for the AB stacking stays similar to those 
for the AA case, i.e., two pairs of the degenerate orbitals are formed. For the 
AA$^{\prime}$ stacking all four product orbitals HOMOs (LUMOs) are separated by the energy gap.

The spatial orbital overlap $J_{i,j}^{H-H}$, regardless of the observed peculiarities of 
orbital mixing as lattice arrangement is changed, is being affected mostly by rearrangements 
of the $\pi$ clouds from their superposition in AA stacking. However, an analysis of 
interaction of the occupied/unoccupied orbitals between fragments $J_{i,j}^{H-L}$ has shown 
a distinct behavior. We observed barely present overlap between those orbitals in 
the AA pattern ($J_{i,j}^{H-L}$= $10^{-5}$ eV), while it appears for the AB stacking to be 
0.169 eV and increases even further up to 0.191 eV for the AA$^{\prime}$ stacking. Such 
progress is consistent with improvement of the orbital interactions $\Delta E_{oi}$ and 
with enhancement of the attractive interactions of the electrostatic nature (see $\Delta 
V_{el}$ in Table ~\ref{tab:table1}) as stacking pattern is changed from AA
to AB. Both these terms, $\Delta E_{oi}$ and $\Delta V_{el}$, lowers the bonding energy 
$\Delta E^0$ and their contribution compensate the growing Pauli repulsion between fragments.

Therefore, the efficiency of the acceptor-donor interactions, defined by $\Delta E_{oi}$, 
is found to be several times (at least five) weaker for the AA stacking in comparison 
to that for the AB. This behavior offers an interpretation of the interlayer decoupling 
observed in the experiments for the AA stacking in twisted bilayer graphene 
\cite{hass,naturp,varchon,reina,warner}, against the efficient coupling known for the 
AB stacking. According to our findings the interlayer decoupling in AA staking must be 
caused by suppression of the orbital overlap of the occupied, $\pi$, and unoccupied, 
$\pi^*$, orbitals between the fragments, i.e. inter-band (HOMO-LUMO) interaction, 
while variation of the $\pi$-$\pi$ overlap is found to have a insignificant affect 
on the interaction between fragments. This conclusion is, in fact, in contrast to 
the commonly accepted opinion that the electronic coupling between the stacked graphene 
layers, which is capable to change linear Dirac cones dispersion to parabolic one,
originates from intra-band interlayer interactions \cite{mcdonald,mccan}. We should 
also note that along with the intra- and inter- band interactions the effect of 
orbital polarization is one of the significant contribution into the bonding interactions.
 
\section{Interlayer distance} 
The observation that the interlayer repulsion characterized by $\Delta E_{p}$ is being 
stronger for the AB stacking in comparison to that for AA was beyond our expectations 
because the Pauli repulsion is presumed to dominate when one structure of closed 
electron shell is placed exactly on top of the other (such as the coordinates in $x-y$ 
directions coincide as presented in Fig.~\ref{fig:fig1} (a)). The Pauli repulsion has 
the exponential dependence on the separation distance and an increase in $d_{eq}$ for 
the AA case must cause a significant suppression of the repulsive forces. To understand 
this, we considered the deviation of the bonding energy and its components with the change 
in the interlayer distance for the AA stacking [Fig.~\ref{fig:fig1} (b)]. The negative sign 
is for the attractive interactions. The electrostatic interactions $\Delta V_{el}$ and the 
orbital interactions $\Delta E_{oi}$ are both attractive in nature and their 
magnitudes are reduced with increasing distance between the fragments ($\Delta V_{el}$ 
can have a positive magnitude only for short interlayer distance when the 
nuclear repulsion dominates over other attractive terms). The contribution of the 
repulsive forces collected within $\Delta E_{p}$ into the total bonding energy 
displays its domination over the attractive terms only for a short interlayer 
distance, so the $\Delta E^0$ remains positive (repulsive) up to a distance of 
3.2 \AA. The bonding energy $\Delta E^0$ reaches its minimum at $d_{eq}\sim$3.67 \AA\ 
($\Delta E^0\sim -1.47$ eV) which is still energetically far from the 
ground state of AB stacking, where $\Delta E^0= -1.93$ eV. 
 
Among all the interaction terms the Pauli repulsion and the bonding interactions are points of 
particular interest. According to the Pauli principle the valence electron from one fragment is
not supposed to penetrate the closed valence shell of the other fragment because the repulsive 
forces expel the charges from the overlap region. The main component within $\Delta E_{p}$ which 
contribute to the repulsion effect comes from the kinetic energy while the potential energy part  
is attractive \cite{bick}. It is noticeable in Fig.~\ref{fig:fig1} (b) that as the fragments being
separated at a distance beyond the value of 4.0 \AA\ the Pauli repulsion becomes negligible.  
However, for the interlayer distance 3.4 \AA\ which is typical for the AB stacking, the Pauli 
repulsion is still strong as $\Delta E_{p}$=4.33 eV while for the AB stacking pattern it was  
3.38 eV. Therefore, if we compare the AA and AB stacking for the same interlayer distance 3.4 \AA, 
the Pauli repulsion is larger by almost 1.0 eV for the AA stacking, in agreement with our 
expectations, as stated above. 
 
\begin{figure*} 
\includegraphics[scale=0.70]{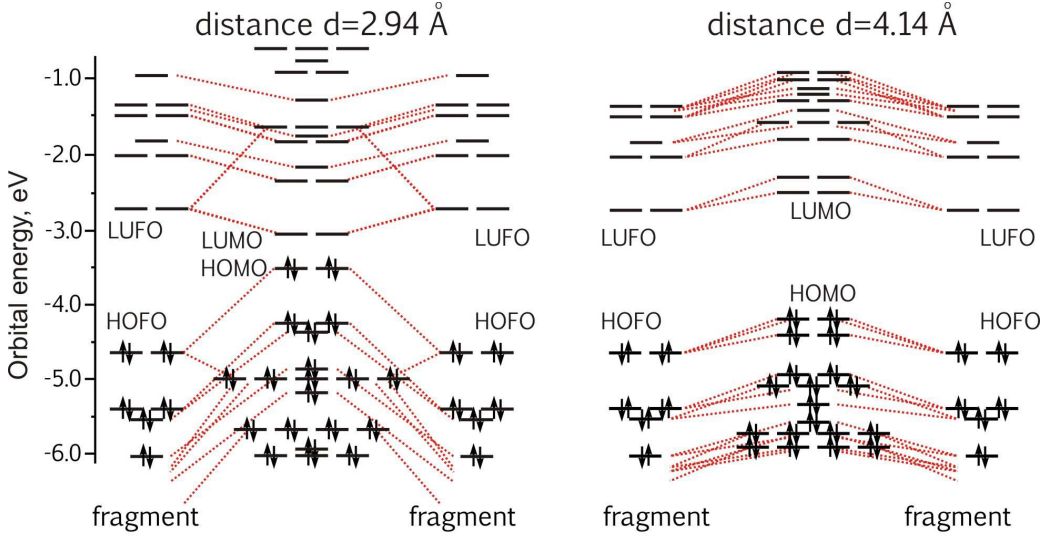} 
\caption{Energetics of the $\pi$ and $\pi^*$ orbitals defining the  
HOMO-LUMO gap in the separated fragments and their splitting as fragments 
being stacked in the AA lattice arrangement.} 
\label{fig:fig3} 
\end{figure*}

The orbital interaction energy $\Delta E_{oi}$ belongs to the attractive forces and 
its value reflects the efficiency of the donor-acceptor charge transfer between 
fragments which is controlled by the interlayer orbital overlap together with the 
Pauli repulsion and orbital polarization. Suppression of the Pauli repulsion with 
growing interlayer distance tends to increase the orbital interaction energy 
$\Delta E_{oi}$. However, a simultaneous reduction of the spatial orbital overlap 
generally leads to diminishing of $\Delta E_{oi}$, i.e., to a reduction of the 
charge transfer between the fragments. As a result, $\Delta E_{oi}$ being strongly 
attractive (negative sign) at short distances almost vanishes as the distance reaches 
the value of $d\simeq$ 3.3 \AA\ while after $d\simeq$3.9 \AA\ it turns repulsive with 
a positive sign. It was noticed that the composition of the molecular orbitals near 
the HOMO-LUMO gap is not changed with distance as it is shown in Fig.~\ref{fig:fig3}.
For example, if the HOMO was formed by the interaction of selective HOFO orbitals provided 
by each fragment it remains of the same composition on a whole range of the distance 
while just become shifted in energy. Therefore, we can conclude that the orbital 
polarization term brings no contribution in deviation of $\Delta E_{oi}$ with distance. 
 
To separate the orbital interactions from other forces, we calculated the spatial 
overlap integral between the fragments $J_{i,j}^{H-H}$ and $J_{i,j}^{H-L}$. The 
degradation of $J_{i,j}^{H-H}$ upon increase of the interlayer distance along with 
the size of the HOMO-LUMO gap ($E_{gap}$) are presented in Table~\ref{tab:table2}. 
Because a rise in interlayer separation $d$ induces a suppression of the orbital 
overlap, in particular of the overlap matrix $S_{i,j}$, the charge transfer integral 
$J_{i,j}^{H-H}$ is also being reduced. The same gradual reduction is observed for 
the $J_{i,j}^{H-L}$ overlap (from 5$\times 10^{-5}$ to 6$\times 10^{-6}$ eV). 
 
\begin{table} 
\caption{\label{tab:table2} The spatial overlap integral 
$J_{i,j}^{H-H}$ and HOMO-LUMO gap $\Delta E_{gap}$ calculated for the AA-stacking pattern  
as the flake separation $d$ gradually increases. All values are in eV.} 
\begin{tabular}{l|c|c|c|c|c|c|c|c} 
\hline 
$d$ & 2.94 & 3.14 & 3.34 & 3.40 & 3.54 & 3.74 & 3.94 & 4.14\\ 
\hline 
$J_{i,j}^{H-H}$ & -1.21 & -0.91 & -0.68 & -0.63 & -0.51 & -0.38 & -0.28 & -0.21\\ 
$\Delta E_{gap}$ & 0.45 & 0.84 & 1.13 & 1.20 & 1.34 & 1.50 & 1.61 & 1.69\\ 
\hline 
\end{tabular} 
\end{table} 
 
As one increases the interlayer distance $d$ the HOMO-LUMO gap grows in contrast 
to the diminishing $J_{i,j}$ and its enhancement is directly connected to the orbital 
interaction between the fragments. We presented in Fig.~\ref{fig:fig3} the energy diagram 
for the energetics of the $\pi$ and $\pi^*$ orbitals near the HOMO-LUMO gap for the 
case of separated fragments and their orbital splitting/mixing after perturbation. To 
find the orbital energy change by the fragment interaction we used the expression derived 
within the H\"uckel approximation for the description of splitting of the $\pi$ orbitals 
belonging to different fragments after inclusion of the interactions
\begin{equation} 
E_{1}\approx e_{0}+H_{12}-(e_{0}+H_{12})S_{12} 
\label{eq:two} 
\end{equation} 
\begin{equation} 
E_{2}\approx e_{0}-H_{12}+(e_{0}-H_{12})S_{12} 
\label{eq:three} 
\end{equation} 
where $e_{0}$ and $E_{1,2}$ are the molecular $\pi$-orbital energies before (for  
identical graphene flakes $e_{1}=e_{2}=e_{0}$) and after perturbation, respectively.  
$S_{12}$ is the orbital overlap between the fragments and  
$H_{12}$ is the intrinsic interaction integral that is a combination of the energy terms 
responsible for electron-electron interactions and particularly  
the contribution of its repulsive part to the orbital energies $E_{1(2)}$. 
 
As fragments are brought together to a distance of 2.94 \AA\ (see Fig.~\ref{fig:fig3})  
which is much shorter than the equilibrium separation ($d_{eq}=3.67$ \AA), 
the repulsive force dominates over the attractive interaction. Therefore, since the 
electrons are expelled by repulsion from the overlap region, within the H\"uckel approximation  
the overlap matrix $S_{12}$ is treated as zero or would possess a negative value so that
for simplicity, we can ignore the contribution from $(e_{0}\pm H_{12})S_{12}$ to the orbital 
energy $E_{1(2)}$. Besides the effect of repulsion on $(e_{0}\pm H_{12})S_{12}$ term,  
the strong repulsion between the fragments causes the increment of the $H_{12}$ 
and therefore, closer the fragments are to each other the larger the splitting of their  
$\pi$ orbitals. The significant $\pi-\pi$ splitting leads to shift of the LUMO  
and HOMO orbitals close to each other thus causing a suppression of the HOMO-LUMO gap. 
However, for a large interlayer distance of 4.14 \AA, the value of $H_{12}$  
diminishes because of the suppression of the Pauli repulsion, while the overlap matrix 
would have positive values. The charge exchange between the fragments is allowed which
decreases the splitting $|E_{1}-E_{2}|$ for the $\pi$ orbitals and in turn enlarges the 
HOMO-LUMO gap ($\Delta E_{gap}$). 
 
In this section we considered the basics of the orbital interactions between two  
graphene fragments being under the control of the separation distance. It was gathered that 
there are two main components affecting the efficiency of the orbital interaction energy 
by varying the distance: the Pauli repulsion and the spatial orbital overlap, while no 
contribution from the orbital polarization was observed. The spatial orbital overlap 
decreases with increasing distance and so is the Pauli repulsion. These changes have the
opposite influence on the orbital interaction energy $\Delta E_{oi}$, but since the 
decrease of the spatial orbital overlap is faster than the Pauli repulsion, $\Delta E_{oi}$ 
is generally reduced with induced flake separation. 
 
\section{Flake rotation} 
For two graphene layers stacked together, the rotational misorientation creates the 
Moir\'e pattern which is a periodic pattern manifesting itself through the spots 
where the superposition of the lattices is preserved, i.e., the AA stacking 
\cite{latil,shall,lopes}. The rest of the surface (the inter-spot regions) which are 
of a larger amount of the surface, possesses the stacking order similar or close to 
that for the AB arrangement (see Ref.~\cite{camp} for images of various rotation structures). 
As the rotation angle increases, the percentage of AA-spotted areas grows while the size 
of the spots and the inter-spot areas shrink. Already for the angle above 20$^{\circ}$ 
the spots with ideal AA stacking vanish completely, leaving the mixed and weakly defined 
interlayer lattice order. Therefore, the experimentally observed lattice misorientation 
induced by rotation angle of 3-5$^{\circ}$ (see Refs.\cite{varchon,naturp,hass,reina,warner})
is characterized by the well defined the AA-spots and of a high coverage of the AB stacking. 
 
Graphene flakes of finite size are considered in this paper as a model system to represent 
the spots characterized by the lattice superposition (the AA-spots) in misoriented bilayer 
graphene. Therefore, by rotating one flake with respect to the other (the rotation axis is 
placed at the flake center) we basically recreate the modification of the shape of the 
AA-spots affected by the rotation angle. The main disadvantage of describing misoriented 
bilayer graphene by finite size flakes is to underestimate the contribution of the AB 
stacking areas to the electronic properties of the bilayer system. In the model system of 
rotated flakes the contribution of the inter-sport regions with AB stacking would depend 
on the flake size and therefore for small flakes would be negligible. 
 
In a system of the AA stacked graphene fragments, the interacting $\pi$ orbitals between 
the fragments are perfectly orthogonal. The layer rotation observed in the experiments  
\cite{hass,naturp,varchon,reina,warner} breaks that orthogonality, thereby modifying the 
balance of the attractive and repulsive forces between the layers. We simulated the effect 
of misorientation in the model system of two graphene flakes presented in Fig.~\ref{fig:fig1} 
(a) with the rotation axis placed at the flake center. In a system of two flakes stacked 
in the AA pattern and separated by a distance of $d$=3.4 \AA\, which is the equilibrium 
distance for the AB-stacking, we rotate one flake relative to the other for further 
elaboration of the interaction parameters. Our simulation results for the bonding energy 
$\Delta E^0$ between layers and its components, overlap integrals such as $J_{i,j}^{H-H}$ 
and $J_{i,j}^{H-L}$, and the HOMO-LUMO gap ($\Delta E_{gap}$) are displayed in 
Table ~\ref{tab:table3}. The results indicate that the size of the HOMO-LUMO gap ($\Delta 
E_{gap}$) grows as the bonding interaction between layers is improved with rotation. 
 
\begin{table} 
\caption{\label{tab:table3} 
The interlayer interactions in the system of two twisted flakes separated by a distance 
$d$=3.4 \AA. For the spatial overlap integral $J_{i,j}^{H-H}$, its modulus has been 
considered to avoid confusions of the sign change. $\Delta E_{gap}$ is the HOMO-LUMO gap 
in the system of two stacked flakes. For a better comparison we repeated the results for 
the graphene flakes stacked in the AB pattern at the equilibrium distance $d_{eq}$=3.4 \AA. 
All values are in eV} 
\begin{tabular}{l|c|c|c|c|c|c|c} 
\hline 
& $\Delta E_{gap}$& $|J_{i,j}^{H-H}|$ & $|J_{i,j}^{H-L}|$& $\Delta E_{oi}$ & $\Delta V_{el}$  
& $\Delta E_{p}$ & $\Delta E^0$ \\ 
\hline 
0.0 & 1.203 & 0.627 & 1.6$\times 10^{-5}$& -0.248 & -1.423 & 4.334 & -1.195 \\ 
1.5 & 1.209 & 0.604 & 2.5$\times 10^{-4}$& -0.253 & -1.421 & 4.325 & -1.207 \\ 
3.0 & 1.226 & 0.544 & 6.5$\times 10^{-4}$& -0.268 & -1.416 & 4.300 & -1.242 \\ 
6.0 & 1.287 & 0.563 & 6.5$\times 10^{-4}$& -0.320 & -1.399 & 4.205 & -1.368 \\ 
9.0 & 1.365 & 0.305 & 8.9$\times 10^{-4}$& -0.384 & -1.376 & 4.081 & -1.530 \\ 
12.0 & 1.436 & 0.116 & 2.3$\times 10^{-4}$& -0.443 & -1.356 & 3.963 & -1.682 \\ 
15.0 & 1.478 & 0.131 & 6.5$\times 10^{-4}$& -0.482 & -1.342 & 3.876 & -1.790\\ 
18.0 & 1.492 & 0.061 & 6.9$\times 10^{-4}$& -0.494 & -1.335 & 3.831 & -1.837\\ 
20.0 & 1.489 & 0.176 & 3.9$\times 10^{-4}$& -0.486 & -1.333 & 3.823 & -1.832\\ 
22.0 & 1.485 & 0.318 & 1.1$\times 10^{-4}$& -0.475 & -1.332 & 3.826 & -1.816\\ 
24.0 & 1.481 & 0.184 & 5.8$\times 10^{-4}$& -0.453 & -1.333 & 3.839 & -1.779\\ 
26.0 & 1.478 & 0.344 & 3.7$\times 10^{-3}$& -0.439 & -1.333 & 3.851 & -1.753\\ 
28.0 & 1.477 & 0.363 & 2.0$\times 10^{-3}$& -0.423 & -1.334 & 3.862 & -1.724\\ 
30.0 & 1.476 & 0.322 & 2.9$\times 10^{-3}$& -0.417 & -1.335 & 3.870 & -1.713 \\ 
AB  & 1.568 & 0.229 & 0.169 &-0.482 & -1.184 & 3.388 & -1.931 \\ 
\hline 
\end{tabular} 
\end{table} 
 
When two lattices of different flakes are superposed in the space, the electronic clouds 
of their $\pi$ orbitals are also superposed giving the maximum magnitude of the overlap 
integral $J_{i,j}^{H-H}$ which is being suppressed with the flake rotation because of 
misorientation of those $\pi$ clouds. Therefore, the spatial overlap integral $J_{i,j}^{H-H}$ 
reaches its minimum as the rotation angle reaches the value of $\theta\simeq$ 18$^{\circ}$  
which is being a result of significant disarrangement of the $\pi$ orbitals 
within the overlapping region from their superposed position. Another overlap integral 
accounting for inter-band interactions, $J_{i,j}^{H-L}$ (between HOMO and LUMO orbitals 
belonging to the different fragments), has shown the opposite behavior to that of 
$J_{i,j}^{H-H}$, i.e., with its minimum for the superposed case while growing with the 
flake rotation. However, even when $J_{i,j}^{H-L}$ reaches the maximum at $\theta=24^{\circ}$-30$^{\circ}$, 
its magnitude is still much lower than that found for the AB-stacking. For the angle in the 
range of $\theta=24^{\circ}$-30$^{\circ}$, both integrals ($J_{i,j}^{H-H}$ and $J_{i,j}^{H-L}$) 
deviate insignificantly. With breaking of the orthogonality of the $\pi$ orbitals as the 
flake is rotated, the Pauli repulsion is also being suppressed by $\sim$ 0.5 eV when its 
reaches its minimum at $\theta\simeq$ 18$^{\circ}$. The reduced value of 3.831 eV for 
$\theta\simeq$ 18$^{\circ}$ becomes much closer to that for the AB stacking  
($\Delta E_{p}$=3.388 eV). 
 
Moreover, the orbital interaction energy $\Delta E_{oi}$ grows with flake rotation as the 
inter-band interactions reflected by the $J_{i,j}^{H-L}$ improve ($J_{i,j}^{H-L}$ increases 
up to several orders of magnitude) along with the fast reduction of the interlayer repulsion. 
The coincidence of the minimum of $\Delta E_{oi}$ with the minimum value of the Pauli repulsion 
achieved for the $\theta\simeq$ 18$^{\circ}$ is a clear evidence that the efficiency of the 
orbital interactions is being under direct control of the repulsive forces. 

Therefore, because the layer rotation significantly suppresses the interlayer repulsion 
which in turn improves the orbital interaction, regardless of the deviation of the overlap 
integrals, the total bonding energy $\Delta E^0$ is lowered with rotation and its magnitude 
reaches its minimum also at $\theta\simeq$ 18$^{\circ}$. In fact, the magnitude of the total 
bonding energy found for the angle $\theta\simeq$ 18$^{\circ}$ ($\Delta E^0$=$-$1.837 eV) is 
comparable to that for the AB stacking ($\Delta E^0$=$-$1.931 eV in Table ~\ref{tab:table1}).  
Above the rotation angle 30$^{\circ}$, some fluctuations for all the terms occur  
while closer to angle 60$^{\circ}$ for which the conditions for the lattice  
superposition between two layers reappear, i.e., all the terms have the same values as 
for the angle 0$^{\circ}$. Basically, the dependence in the range of the rotation angle 
from 30$^{\circ}$ to 60$^{\circ}$ is displayed in reverse order to  
that from 0$^{\circ}$ to 30$^{\circ}$. 
 
Our main conclusion is that the repulsion appeared as result of the interaction of two  
systems of closed electron shells whose lattices are superposed, is the central force 
controlling the efficiency of the interlayer orbital interactions. As the flake rotation 
breaks the lattice superposition, the suppression of the repulsion induces an improvement 
of orbital interactions (such as the electronic coupling) thereby lowering the bonding 
energy $\Delta E^0$. Because the orbital interaction between the flakes depends on the  
flake rotation, the equilibrium distance $d_{eq}$ also fluctuates with rotation.  
The modification of the equilibrium distance $d_{eq}$ follows the dependence  
observed for the bonding energy $\Delta E^0$ which is controlled by the Pauli repulsion.  
The maximum equilibrium distance $d_{eq}$=3.67 \AA\ is obtained for the AA stacking which 
is suppressed down to $d_{eq}$=3.43 \AA\ as the flakes are at an angle 18$^{\circ}$ 
that brings the system to the lowest energy state achievable with rotation.  
If the rotational angle grows further beyond 18$^{\circ}$, the magnitude of $d_{eq}$ 
enhances again and for $\theta\simeq$ 30$^{\circ}$ its value is 3.54 \AA. 
 
\section{Discussion and summary} 
Two graphene layers stacked in the AA pattern which is characterized by the lattice 
superposition between the layers is the most unstable configuration in the bilayer 
geometry. The instability appears as a result of strong interlayer repulsion  
induced by the interaction of the filled orthogonal $\pi$ orbitals within the  
overlap region. Therefore, a disruption of the lattice superposition lowers the  
total bonding energy and therefore, leads to an enhancement of the system stability. 
The AB stacking is the most successful scheme to suppress the Pauli repulsion because 
it induces the maximum mis-orientation in the interlayer lattice order and therefore, 
the AB stacking appear to be the ground state of the system characterized by the  
strongest bonding interactions between the layers. However, an alternative way to 
induce the lattice mis-orientation from that superposed in the AA stacking
and thereby, to transfer the system to the lower energy state is the layer rotation. 
It should be noted that any modification of the lattice order different from the 
AB-stacking would be metastable (such as AA$^{\prime}$) because AB stacking is 
the ground state of the system.  
 
For adjacent graphene flakes of finite size, the rotation of one flake relative 
to the other induces a fast reduction of the repulsive part ($\Delta E_{p}$) and 
an increase of the attractive forces (orbital interaction energy $\Delta E_{oi}$) 
such that both these tendencies lower the total bonding energy between the layers 
$\Delta E^0$. The lowest bonding energy $\Delta E^0$ is achieved for the rotation 
angle of $\theta \simeq$ 18$^{\circ}$. This state is still metastable but with the 
lowest value of the bonding energy among all the rotation angles, and its value 
correlates with that for the AB-stacking being the ground state of bilayer graphene. 
However, as it was already noted above, the description of the adjacent graphene 
layers by the model system of finite flakes has crucial disadvantage caused by 
the underestimation of the inter-spot areas of the AB-stacking into the interlayer 
repulsion. As we switch to the twisted bilayer graphene of the infinite size, 
the contribution of large areas of AB stacking, which was largely neglected in the 
flake system, should be considered. For a small rotation angle the percentage of the  
AA spotted areas is large which decreases with angle enhancement. Thus, for an angle 
altered from 10$^{\circ}$ to 12$^{\circ}$, a decrease of 2.5 \% of AA staking is observed 
\cite{camp}. Obviously, since the larger inter-spot areas of AB-stacking is 
observed for small rotation angles 2$^{\circ}$-5$^{\circ}$ (see Ref.\cite{camp} for the images 
of various rotation structures), we would expect that the rotation angle of much smaller 
magnitude than that for the finite systems might be required ($\theta <$ 18$^{\circ}$) 
to bring the stacked graphene layers to the metastable state with the lowest energy. 
  
However, additionally to rotation there is another way to make 
the system of the flakes stacked in the AA arrangement more stable which is  
to raise the interlayer distance when the bonding energy is lowered again
due to suppression of the Pauli repulsion between the layers (see Fig.~\ref{fig:fig1} (b)).  
In case of misoriented bilayer graphene exhibiting the Moir\'e pattern the distribution  
of the repulsive forces would be non-uniform as the lattice order is not the same 
in different areas, i.e. a maximum force pushing apart two lattices would originate 
at the AA-spots of the lattice superposition. Moreover, another interesting distinction  
between the system of adjacent flakes and bilayer graphene of infinite size is alteration 
of its rigidity. Recalling that the free standing graphene is subjected to rippling 
of its surface \cite{ripples}, the lower rigidity of the graphene layers than that 
of flakes is anticipated. Therefore, we expect that a strong Pauli repulsion  
which is pronounced locally at the center of the AA-spots might not able to modify 
the interlayer separation throughout the whole system because of large areas of 
AB-stacking, but would rather induce a local lattice distortion forming the bump 
on the surface with its highest point at the center of the AA-spot \cite{our}. To 
simulate this effect, the flake of larger size containing the bigger areas of AB 
stacking have been examined and already for that system the generation of the bump 
as high as 0.2 \AA\ was observed. However, the bump's height may be enhanced for 
an infinite system due to better efficiency of the attractive interactions in the  
inter-spot areas and lower rigidity of the layers. In fact, the appearance of bumps can 
explain the brightening of the AA-spots observed in the STM images of the twisted bilayer 
graphene \cite{varchon,hass,naturp}. 
 
The final point we wish to make is about the interlayer coupling which 
is found to be a function of the rotation angle and interlayer distance.
As fragments being stacked, there are two types of $\pi$ interactions occurs,
such as perturbation of the occupied/occupied orbitals ($\pi$-$\pi$ interaction) 
and interaction of the occupied/unoccupied orbitals ($\pi$-$\pi^*$ interaction). 
According to the theoretical models developed to describe the behavior of the 
$\pi$ bands in bilayer graphene, the modification 
of the linear dispersion of the Dirac cones to a parabolic one
has been simulated by inclusion only of the $\pi$-$\pi$ orbital interaction 
\cite{mcdonald,mccan}. However, we found that although 
the intra-bands interactions plays an important role, 
but particularly the inter-bands part must be introduced in the model to 
account for the decoupling effect arising in the AA stacked graphene layers.
The orbital interaction energy  $\Delta E_{oi}$ is therefore suppressed 
at least by five times as the lattice arrangement was changed from AB to AA due to 
vanishing of the $\pi$-$\pi^*$ interactions between fragments 
that is reflected by a drastic reduction of the spatial overlap integral $J_{i,j}^{H-L}$.
 
The work was supported by the Canada Research Chairs program.

\end{document}